\documentclass[12pt]{article} 
\usepackage{amssymb}
\usepackage{graphicx}
\begin{document} 
\begin{center}
          {\Large \bf Hadron structure and elastic scattering} 

\vspace{0.5cm}                   
{\bf I.M. Dremin\footnote{Email: dremin@lpi.ru}}

\vspace{0.5cm}              
          {\it Lebedev Physical Institute, Moscow 119991, Russia}

\end{center}

{\bf Prelude.}

It was in September 1954 when I met Pomeranchuk. He lectured Theory of
Relativity. Next year he became a tutor of our "theoretical group". In some
time he proposed me to learn the book of Fl{\" u}gge on nuclear physics.
After struggling with difficult subject written in German I asked him some 
questions. He liked it and one by one proposed several problems to be solved.
In parallel, he insisted on passing through the series of famous Landau exams.
It was a good school. Its lessons I described in the book of reminiscences
about Pomeranchuk. As a supervisor, he advised my diploma paper to be published 
in JETP Letters
and recommended me to Prof. Tamm as a PhD student. Soon I proposed the one-pion
exchange model which was later extended to multiperipheral and multireggeon
models. Pomeranchuk got interested in it and asked me to come for discussion.
He was deeply interested in properties of hadron collisions.
Our contacts lasted till his death.

In this paper I describe some new findings about elastic scattering of
hadrons studied now up to LHC energies. I briefly reviewed this at the 
Pomeranchuk centennial seminar at ITEP. This would be extremely interesting 
subject for him. Let me just mention Pomeron and famous Pomeranchuk
theorem to remind you his basic contributions in this field. I dedicate 
the paper to the memory of my teacher Isaak Pomeranchuk.

\begin{abstract}
When colliding, the high energy hadrons can either produce new particles or
scatter elastically without change of their quantum numbers and other particles 
produced.  Namely elastic scatterings of hadrons are considered in this paper.
The general machinery of their theoretical treatment is described. Some new 
experimental data are presented and confronted to phenomenological
approaches. The internal structure of hadrons is the main subject of these 
studies. Its impact on properties of their interactions is reviewed. 
It is shown that protons become larger and darker with increase of their 
collision energy and reveal some substructure. The violation of the geometric
scaling in the diffraction cone and new problems of description of differential
cross sections outside it are described. 
\end{abstract}

\section{Introduction}

Hadron interactions are strong and, in principle, should be described
by quantum chromodynamics (QCD). However, experimental data show that their 
main features originate from the non-perturbative sector of QCD. Only the 
comparatively rare processes with high transferred momenta can be treated 
theoretically rather successfully by the perturbative methods due to the 
well-known property of the asymptotical freedom of QCD. Thus, in absence of
methods for rigorous solution of QCD equations, our understanding 
of the dynamics of the main bulk of strong interactions is severely limited 
by the model building or some rare rigorous relations. In fact, our approach 
to high energy hadronic processes at present is at best still in its infancy.

From experiment we have learned, at least, about five subregions of the elastic 
scattering differential cross section. Here, we discuss only two of them: the 
diffraction cone and the Orear regime. The diffraction cone at small angles 
reminds the
quasiclassical effects with Gaussian decrease in angles. The region at larger 
angles with exponential decrease of the cross section called Orear region 
became noticeable only at energies of colliding particles above several GeV. 
It persists till present LHC energies of 7 and 8 TeV. 

The extended review was earlier published in \cite{ufnel}. The sections 2 and 3
are the abbreviated versions of its corresponding parts. Others present some
results obtained later.

\section{The main relations}

The measurement of the differential cross section is the only source of the 
experimental information about their elastic scattering. Herefrom,
the main characteristics of hadron interactions directly related to the elastic 
scattering amplitude such as the total cross section, the elastic scattering
cross section, the ratio of the real to imaginary part of the amplitude, the 
slope of the diffraction cone etc are obtained. The first two of them
are functions of the total energy only, while others depend on two variables -
the total energy and the transferred momentum (or the scattering angle).

The dimensionless elastic scattering amplitude $A$ defines the 
differential cross section in a following way:
\begin{equation}
\frac {d\sigma (s)}{dt}=\frac {1}{16\pi s^2}\vert A\vert ^2=
\frac {1}{16\pi s^2}({\rm Im}A(s,t))^2(1+\rho ^2(s,t)).
\label{dsel}
\end{equation}
Here, the ratio of the real and imaginary parts of the amplitude has been 
defined
\begin{equation}
\rho (s,t) = \frac {{\rm Re}A(s,t) }{{\rm Im}A(s,t)}.
\label{rho}
\end{equation}
In what follows, we consider the very high energy processes. Therefore,
the masses of the colliding particles may be neglected, and one uses the 
expression $s=4E^2\approx 4p^2$, where $E$ and $p$ are the energy and the
momentum in the center of mass system. The four-momentum transfer squared is
\begin{equation}
-t=2p^2(1-\cos \theta )\approx p^2\theta ^2\approx p_t^2 \;\;\;\;\; 
(\theta \ll 1)
\label{trans}
\end{equation}
with $\theta $ denoting the scattering angle in the center of mass system
and $p_t$ the transverse momentum. 

The elastic scattering cross section is given by the integral of the
differential cross section (\ref{dsel}) over all transferred 
momenta:
\begin{equation}
\sigma _{el}(s)=\int_{t_{min}}^{0} dt \frac {d\sigma (s)}{dt}.
\label{sel}
\end{equation}
              
The total cross section $\sigma _t$ is related by the optical theorem with 
the imaginary part of the forward scattering amplitude at high energy $s$ as
\begin{equation}
\sigma _t(s)= \frac {{\rm Im}A(p,\theta =0) }{s}.
\label{sigt}
\end{equation}     

Elastically scattered hadrons escape from the interaction region declining
mostly at quite small angles within the 
so-called diffraction cone\footnote {The tiny region of the interference of the 
Coulomb and nuclear amplitudes at extremely small angles does not contribute 
to the elastic scattering cross section and we discard it.}. Therefore 
the main attention has been paid to this 
region. As known from experiment, the diffraction peak has a Gaussian shape in 
the scattering angles or exponentially decreasing as the function of the 
transferred momentum squared:
\begin{equation}\
\frac {d\sigma }{dt}/\left( \frac {d\sigma }{dt}\right )_{t=0}=e^{Bt}\approx 
e^{-Bp^2\theta ^2}.
\label{diff}
\end{equation}
In view of the relations (\ref{sel}), (\ref{sigt}), (\ref{diff}), any successful 
theoretical description of the differential distribution must succeed in fits
of the energy dependence of the total and elastic cross sections as well.
 
The diffraction cone slope $B$ is given by
\begin{equation}
B(s,t)\approx \frac {d}{dt} \left [\ln \frac {d\sigma (s,t)}{dt}\right ].
\label{bst}
\end{equation}
In experiment, the slope $B$ depends slightly on $t$ at the given energy $s$. 
E.g., at the LHC, its value changes by about 
10$\%$ within the cone for $\vert \Delta t\vert \approx 0.3$ GeV$^2$. We neglect
it in a first approximation.

The normalization factor in Eq. (\ref{diff}) is
\begin{equation}
 \left (\frac {d\sigma }{dt}\right )_{t=0}=\frac {\sigma _t^2(s)(1+\rho _0^2(s))}
{16\pi }, 
\label{diff0}
\end{equation}
where $\rho _0=\rho (s,0)$.
Eq. (\ref{diff0}) follows from Eqs (\ref{dsel}) and  (\ref{sigt}) at $t=0$.

According to the dispersion relations which connect the real and imaginary parts 
of the amplitude and the optical theorem Eq. (\ref{sigt}), the value $\rho _0$ 
may be expressed as an integral
of the total cross section over the whole energy range. In practice $\rho _0$
is mainly sensitive to the local derivative of the total cross section. Then
to a first approximation the result of the dispersion relation may be written
in a form \cite{gmig, sukha, fkol}
\begin{equation}
\rho _0(s)\approx \frac {1}{\sigma _t}\left [\tan \left (\frac {\pi }{2}
\frac {d}{d\ln s }\right )\right ]\sigma _t=
\frac {1}{\sigma _t}\left [ \frac {\pi }{2}\frac {d}{d\ln s }+
\frac {1}{3}\left (\frac {\pi }{2}\right )^3\frac {d^3}{d\ln s^3 }+...\right ]
\sigma _t.
\label{rhodi}
\end{equation}
At high energies $\rho _0(s)$ is mainly determined by the derivative of the 
logarithm of the total cross section with respect to the logarithm of energy.

The bold extension of the first term in this series to non-zero transferred
momenta would look like
\begin{equation}
\rho (s,t)\approx \frac {\pi }{2}
\left [\frac {d\ln {\rm Im}A(s,t) }{d\ln s }-1\right ].
\label{rhodit}
\end{equation}

If one neglects the high-$\vert t\vert $ tail of the differential cross section,
which is several decades lower than the optical point, and integrates in 
Eq. (\ref{sel}) using Eq. (\ref{diff}) with constant $B$, then one 
gets the approximate relation between the total cross section, the elastic cross
section and the slope 
\begin{equation}
\frac {\sigma _t^2(1+\rho _0^2)}{16\pi B\sigma _{el}}\approx 1.
\label{stseb}
\end{equation}

The elastic scattering amplitude must satisfy the general properties of 
analiticity, crossing symmetry and unitarity. The unitarity of the $S$-matrix 
$SS^+$=1 imposes definite requirements on it. In the $s$-channel it looks like 
\begin{eqnarray}
{\rm Im}A(p,\theta )= I_2(p,\theta )+F(p,\theta )= \nonumber  \\
\frac {1}{32\pi ^2}\int \int d\theta _1
d\theta _2\frac {\sin \theta _1\sin \theta _2A(p,\theta _1)A^*(p,\theta _2)}
{\sqrt {[\cos \theta -\cos (\theta _1+\theta _2)] 
[\cos (\theta _1 -\theta _2) -\cos \theta ]}}+F(p,\theta ).
\label{unit}
\end{eqnarray}
The region of integration in (\ref{unit}) is given by the conditions
\begin{equation}
\vert \theta _1 -\theta _2\vert\leq \theta ,       \;\;\;\;\;
\theta \leq \theta _1 +\theta _2 \leq 2\pi -\theta .
\label{integr}
\end{equation}
The integral term represents the two-particle intermediate states of the 
incoming particles. The function $F(p,\theta )$ represents the shadowing 
contribution of the inelastic processes to the elastic scattering amplitude. 
Following \cite{hove} it is called the overlap function. It determines 
the shape of the diffraction peak and is completely non-perturbative.
Only some phenomenological models pretend to describe it. 

In the forward direction $\theta$=0 this relation in combination with the 
optical theorem (\ref{sigt}) reduces to the  general 
statement that the total cross section is the sum of cross sections of elastic 
and inelastic processes:
\begin{equation}
\sigma _t=\sigma _{el}+\sigma _{inel}.
\label{tein}
\end{equation}

The unitarity relation (\ref{unit}) has been successfully used 
\cite{anddre, anddre1, adg, dnec} for the model-independent description of the 
Orear region between the diffraction cone and hard parton scattering which 
became the crucial test for phenomenological models.

Experimentally, all characteristics of elastic scattering are measured as
functions of energy $s$ and transferred momentum $t$. However, it is
appealing to get knowledge about the geometrical structure of scattered
particles and the role of different space regions in the scattering process.
Then one should use the Fourier-Bessel transform to get correspondence between
the transferred momenta and these space regions. The transverse distance between 
the centers of colliding particles called as the impact parameter $\bf b$ 
determines the effective transferred momenta $t$. The amplitudes in the
corresponding representations are related as
\begin{equation}
h(s,b)=\frac {1}{16\pi s}\int _{t_{min}=-s}^0dtA(s,t)J_0(b\sqrt {-t}).
\label{hsb}
\end{equation}
Peripheral collisions
at large $b$ lead to small transferred momenta $\vert t\vert $.

The amplitude $A(s,t)$ may be connected to the eikonal phase 
$\delta (s,\bf b)$ and to the opaqueness (or blackness) $\Omega (s,\bf b)$ 
at the impact parameter $\bf b$ by the Fourier-Bessel transformation
\begin{equation}
A(s,t=-q^2)=\frac {2s}{i}\int d^2be^{i{\bf qb}}(e^{2i\delta (s,\bf b)}-1)=
2is\int d^2be^{i{\bf qb}}(1-e^{-\Omega (s,\bf b)}). 
\label{eik}
\end{equation}

Assuming $\Omega (s,\bf b)$ to be real and using Eq. (\ref{sigt}) one gets
\begin{equation}
\sigma _t=4\pi \int _0^{\infty }(1-e^{-\Omega (s,\bf b)})bdb.
\label{cstb}
\end{equation}
Also
\begin{equation}
\sigma _{el}=2\pi \int _0^{\infty}(1-e^{-\Omega (s,\bf b)})^2bdb,
\label{cselb}
\end{equation}
and
\begin{equation}
B=\frac {\int _0^{\infty }(1-e^{-\Omega (s,\bf b)})b^3db}
{2\int _0^{\infty}(1-e^{-\Omega (s,\bf b)})bdb}.
\label{Bb}
\end{equation}

To apply the inverse transformation one must know the amplitude $A(s,t)$ at all
transferred momenta. Therefore, it is necessary to continue it analytically to 
the unphysical region of $t$ \cite{adac}. This may be done \cite{isla}.
Correspondingly, the mathematically consistent inverse formulae contain, 
in general, the sum of contributions from the physical and unphysical parts of 
the amplitude $A(s,t)$. The amplitude in (\ref{unit}) enters only in the 
physical region. Only this part of its Fourier-Bessel transform is important 
in the unitarity relation for the impact parameter representation as well. It is 
written as
\begin{equation}
{\rm Im}h(s,b)= \vert h(s,b)\vert ^2+F(s,b),
\label{unib}
\end{equation}
where $h(s,b)$ and $F(s,b)$ are obtained by the direct transformation of 
$A(s,t)$ and $F(s,t)$ integrated only over the physical transferred momenta 
from $t_{min}$ to 0.
They show the dependence of the intensity of elastic and inelastic interactions 
on the mutual impact parameter of the colliding particles. The integrals over
all impact parameter values in this relation represent analogously to the
relation (\ref{tein}) the total, elastic and inelastic cross-sections, 
respectively. 

However, the 
accuracy of the unitarity condition in $b$-representation (\ref{unib}) is still 
under discussion (see, e.g., \cite{kkl}) since some corrections due to
unphysical region enter there even though their role may be negligible.

\section{Where do we stand now?}

First, let us discuss what we can say about asymptotic properties of such
fundamental characteristics as the total cross section $\sigma _t$, the elastic
cross section $\sigma _{el}$, the ratio of the real part to the imaginary part 
of the elastic amplitude $\rho $ and the width of diffraction peak $B$
at infinite energies. Then we compare this with some trends of
present experimental data. 

More than half a century ago it was claimed \cite{froi, mart} that 
according to the general principles of the field theory and ideas about
hadron interactions the total cross section can not increase with energy 
faster than $\ln ^2s$. The upper bound was recently improved \cite{mar2}
with the coefficient in front of the logarithm shown to be twice smaller
than in the earlier limit:
\begin{equation}
\sigma _t\leq \frac {\pi }{2m_{\pi }^2}\ln^2(s/s_0),
\label{asymp}
\end{equation} 
where $m_{\pi }$ is a pion mass. 

If estimated at present energies, this bound is still much higher than the 
experimentally measured values of the cross sections with $s_0$=1 GeV$^2$
chosen as a "natural" scale. Therefore it is of the 
functional significance. It forbids extremely fast growth of the total cross 
section exceeding asymptotically the above limits. Both the coefficient
in front of logarithm in (\ref{asymp}) and the constancy of $s_0$ are often
questioned. In particular, some possible dependence of $s_0$ on energy $s$
is proclaimed (see, e.g., \cite{azim}).

The Heisenberg uncertainty relation points out that such a regime favors the 
exponentially bounded space profile of the distribution of matter density 
$D(r)$ in colliding particles of the type $D(r)\propto \exp (-mr)$. Since the 
energy density
is $ED(r)$ and there should be at least one created particle with mass $m$
in the overlap region, then the condition $ED(r)=m$ gives rise to
$r\leq \frac {1}{m}\ln (s/m^2)$ and, consequently, to the functional dependence 
of (\ref{asymp}). 

It was namely Heisenberg who first proposed earlier just such a behavior
of total cross sections \cite{heis}. He considered the pion production
processes in proton-proton collisions as a shock wave problem governed by some
non-linear field-theoretical equations.

To study asymptotics, some theoretical arguments based on general principles
of field theory and analogy of strong interactions to massive quantum 
electrodynamics \cite{chen} were promoted. The property that the limits
$s\rightarrow \infty $ and $M\rightarrow 0$ (where $M$ denotes the photon 
mass) commute has been used \cite{cw2}, i.e. the asymptotics of strong 
interactions coincides with the massless limit of quantum electrodynamics. 
These studies led to the general geometrical picture of the two hadrons 
interacting as Lorentz-contracted black disks at asymptotically high energies 
(see also the review paper \cite{mbrc}). In what follows, we discuss
some other possibilities as well. However, as a starting point for further
reference, we describe the predictions of this proposal.

The main conclusions are:

1. For black ($\Omega (s,\bf b)\rightarrow \infty $) and logarithmically 
expanding disks with finite radii $R$ ($R=R_0\ln s,\; R_0$=const) one gets from 
(\ref{cstb}) that $\sigma _t$ approaches infinity at asymptotics as
\begin{equation}
\sigma _t(s)=2\pi R^2+O(\ln s); \;\;\;\; R=R_0\ln s; \;\;\;\; R_0={\rm const}.
\label{cst}
\end{equation}

2. The elastic and inelastic processes should contribute on equal footing
\begin{equation}
\frac {\sigma _{el}(s)}{\sigma _t(s)}=
\frac {\sigma _{in}(s)}{\sigma _t(s)}=\frac {1}{2}\mp O(\ln ^{-1}s).
\label{frcs}
\end{equation}
This quantum-mechanical result differs from "intuitive"$\;$ classical
predictions. 

3. The width of the diffraction peak $B^{-1}(s)$ should shrink because
its slope increases as
\begin{equation}
B(s)=\frac {R^2}{4}+O(\ln s) \;\;\;\;\;\;\;\; ({\rm see \;\; also} \;\; \cite{kino}).
\label{wid}
\end{equation}

4. The forward ratio of the real part to the imaginary part of the amplitude
$\rho _0$ must vanish asymptotically as
\begin{equation}
\rho _0=\frac {\pi }{\ln s}+O(\ln ^{-2}s).
\label{rho0}
\end{equation}
This result follows directly from Eq. (\ref{rhodi}) for 
$\sigma _t\propto \ln ^2s$.

5. The differential cross section has the shape reminding the classical 
diffraction of light on the disk
\begin{equation}
\frac {d\sigma }{dt}=\pi R^4\left [\frac {J_1(qR)}{qR}\right ]^2,
\label{bdis}
\end{equation}
where $q^2=-t$.

6. The product of $\sigma _t$ with the value $\gamma $ of $\vert t\vert $ at
which the first dip in the differential elastic cross section occurs is a
constant independent of the energy
\begin{equation}
\gamma \sigma _t=2\pi ^3\beta _1^2+O(\ln ^{-1}s)=35.92 {\rm \; mb}\cdot {\rm GeV}^2,
\label{gams}
\end{equation}
where $\beta _1=1.2197$ is the first zero of $J_1(\beta \pi )$.

These are merely a few conclusions among many model dependent ones.

None of these asymptotical predictions were yet observed in experiment.

Surely, there is another more realistic at present energies
possibility that the black disk model 
is too extreme and the gray fringe always exists. It opens the way to the 
numerous speculations with many new parameters about the particle shape and 
opacity (see the list of references in \cite{ufnel}).

In the Table 1 we show 
the predictions of the gray disk model with the steep rigid edge described by 
the Heaviside step-function and the Gaussian disk model. 
$\Gamma (s,b)$ is the diffraction profile function. 

The slope $B$ is completely determined by the size of the interaction region 
$R$. Other characteristics are sensitive to the blackness of disks $\alpha $.
In particular, the ratio $X$ is proportional to $\alpha $.
The ratio $Z$ plays an important role for fits at larger angles. 
It is inverse proportional to $\alpha $.
The corresponding formulae are given by (\ref{cstb}), 
(\ref{cselb}) and (\ref{Bb}). The black disk limit follows from the gray disk
model at $\alpha =1$. 

\medskip
\begin{table}
\medskip
Table 1. $\;\;$ The gray and Gaussian disks models $\;\; 
(X=\sigma _{el}/\sigma _t, \; Z=4\pi B/\sigma _t) $
\medskip
   
\begin{tabular}{|l|l|l|l|l|l|l|l|}
        \hline
       Model &$1-e^{-\Omega }=\Gamma (s,b)$&$\sigma _t$&
$B$&$X$&$Z$&$X/Z$&$XZ$ \\ \hline
       Gray &$\alpha \theta (R-b);0\leq\alpha<1$&$2\pi \alpha R^2$&
$ R^2/4$&$\alpha /2$&$1/2\alpha$&$\alpha ^2$&$1/4$\\ 
       Gauss &$\alpha e^{-b^2/R^2}; 0\leq \alpha \leq 1$&$2\pi \alpha R^2$
&$R^2/2$&$\alpha /4$&$1/\alpha$&$\alpha ^2/4$&$1/4$ \\
        \hline
    \end{tabular}
\end{table}
The parameter $XZ$ is constant in these models and does not depend on the
nucleon transparency. On the contrary, the parameter $X/Z$ is very sensitive 
to it being proportional to $\alpha ^2$. Therefore, it would be extremely 
instructive to get some knowledge about them from experimental data. 

In the Table 2 we show how the above ratios evolve with energy according to 
experimental data. Most primary entries there except the last two are taken 
from Refs \cite{cyan, chao} with simple recalculation $Z=1/4Y$. The data at 
Tevatron and LHC energies are taken from Refs \cite{am, toteml, totemh}. All 
results are for $pp$-scattering except those at 546 and 1800 GeV for $p\bar p$
processes which should be close to $pp$ at these energies. 

\medskip
\begin{table}
\medskip
Table 2.  $\;\;$ The energy behavior of various characteristics of elastic scattering.
\medskip

    \begin{tabular}{|l|l|l|l|l|l|l|l|l|l|l|l}
        \hline
$\sqrt s$, GeV&2.70&4.11&4.74&6.27&7.62&13.8&62.5&546&1800&7000\\ \hline
   X&0.42&0.28&0.27&0.24&0.22&0.18&0.17&0.21&0.23&0.25 \\  
   Z&0.64&1.02&1.09&1.26&1.34&1.45&1.50&1.20&1.08&1.00 \\  
   X/Z&0.66&0.27&0.25&0.21&0.17&0.16&0.11&0.18&0.21&0.25 \\ 
   XZ &0.27&0.28&0.29&0.30&0.30&0.26&0.25&0.26&0.25&0.25\\  \hline
   
\end{tabular}
\end{table}
The most interesting feature of the experimental results is the minimum of the 
blackness parameter $\alpha $ at the ISR energies.
It is clearly seen in the minima of $X$ and $X/Z$ and in the maximum of $Z$
at $\sqrt s$=62.5 GeV. The steady decrease of ratios $X$ proportional to 
$\alpha $ and $X/Z$ proportional to $\alpha ^2$ till the ISR energies and  
their increase at S$p\bar p$S, Tevatron and LHC energies means that the 
nucleons become more transparent till the ISR energies and more black to 7 TeV.
The same conclusion follows from the behavior of $Z$.
The value of $Z$ approaches fast its limit for the Gaussian distribution of 
matter in the disk. This shows that the above crude models 
are not very bad for qualitative estimates in a first approximation. 

We briefly comment on some of the 
important general trends of high-energy data observed in experiment. 

1. Total cross sections increase with energy. At present energies, the 
power-like approximation is the most preferable one. 

2. The ratio $\sigma _{el}/\sigma _t $ decreases from low energies to those of 
ISR where it becomes approximately equal to 0.17 and then strongly increases 
up to 0.25 at the LHC energies. However, it is still pretty far from the 
asymptotical value 0.5 corresponding to the black disk limit. 

3. The diffraction peak shrinks about twice from energies
about $\sqrt s \approx 6$ GeV where $B \approx 10$ GeV$^{-2}$ to the LHC
energy where $B \approx 20$ GeV$^{-2}$. 

4. A dip and subsequent maximum appear just at the end of the diffraction cone.

5. As regards the behavior of the differential cross section in the function of 
the transverse momentum behind the dip, the $t$-exponential of the 
diffraction peak is replaced, according to experimental data, by the 
$-\sqrt {\vert t\vert }\approx -p_t$-exponential at the intermediate angles:
\begin{equation}
d\sigma /dt\propto e^{-2a\sqrt {\vert t\vert}}, \;\;\; a\approx \sqrt B.
\label{orear}
\end{equation} 
The slope $2a$ in this region increases with energy and it shifts to lower 
$\vert t\vert$. 

6. As a function of energy, the ratio $\rho _0$ increases from negative values
at comparatively low energies, crosses zero in the region of hundreds GeV
and becomes positive at higher energies where it passes through the maximum of
about 0.14 and becomes smaller (about 0.11) at LHC energies.

7. The product $\gamma \sigma _t$ changes from 39.5 mb$\cdot $GeV$^2$ at 
$\sqrt s=6.2$ GeV to 51.9 mb$\cdot $GeV$^2$ at $\sqrt s=7$ TeV
and deviates from the predicted asymptotic value (\ref{gams}).
The total cross section $\sigma _t$ increases faster than $\gamma $ decreases.

From the geometrical point of view the general picture is that protons 
become blacker, edgier and larger (BEL) \cite{rhpv}. We discuss it later.
Thus even though the qualitative trends may be considered as  
satisfactory ones, we are still pretty far from asymptotics even at  
LHC energies. 

\section{LHC data and phenomenology}

Here, we limit ourselves by the latest results of the TOTEM collaboration at 
the highest LHC energies 7 and 8 TeV \cite{toteml, totemh}. The discussion of 
theoretical models is also concentrated near these data.

The total and elastic cross sections at 7 TeV are respectively estimated as 
98.3 mb and 24.8 mb.
The cross section shape in the region of the diffraction cone \cite{toteml} is 
shown in Fig. 1. 
\begin{figure}
\includegraphics[width=\textwidth, height=8cm]{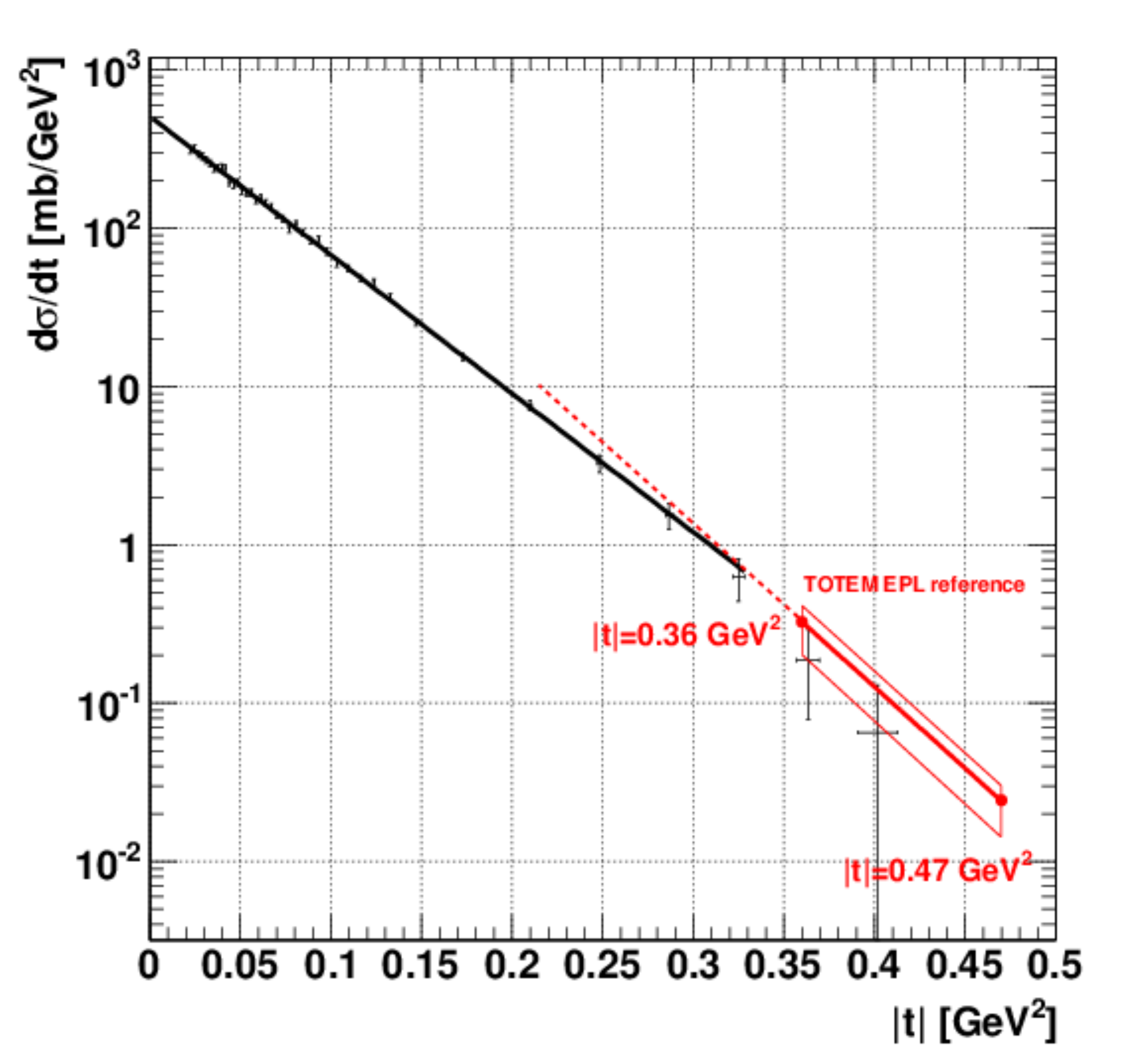}

Fig. 1. The differential cross section of elastic proton-proton scattering at 
$\sqrt s$=7 TeV measured by the TOTEM collaboration  
(Fig. 4 in \cite{totemh}). \\
The region of the diffraction cone with the $\vert t\vert $-exponential 
decrease is shown.
\end{figure}
The $t$-exponential behavior with $B\approx$20.1 GeV$^{-2}$ is 
clearly seen at $\vert t\vert <0.3$ GeV$^2$. The peak steepens at the end of 
the diffraction cone so that
in the $\vert t\vert $ interval of (0.36 -- 0.47) GeV$^2$ its slope becomes
approximately equal to 23.6 GeV$^{-2}$. The results at somewhat larger angles 
\cite{totemh} in the Orear region are presented in Fig. 2. 
\begin{figure}
\centerline{\includegraphics[width=\textwidth, height=8cm]{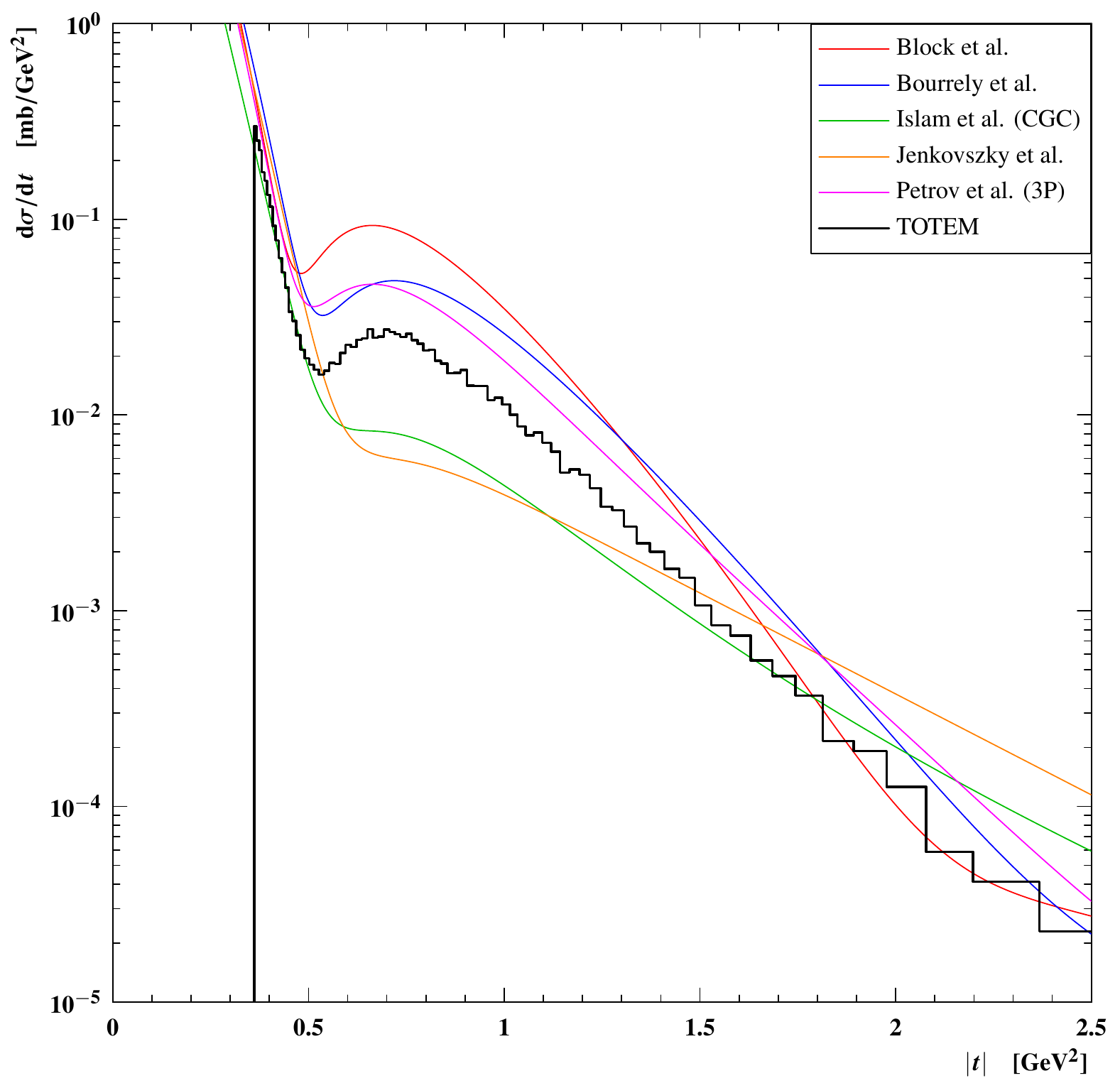}}

Fig. 2. The differential cross section of elastic proton-proton scattering at 
$\sqrt s$=7 TeV measured by the TOTEM collaboration  
(Fig. 4 in \cite{toteml}). \\
The region beyond the diffraction peak is shown. The predictions of five models
are demonstrated. 
\end{figure}
The dip at
$\vert t\vert \approx 0.53$ GeV$^2$ with subsequent maximum at 
$\vert t\vert \approx 0.7$ GeV$^2$ and the $\sqrt {\vert t\vert }$-exponential 
behavior are demonstrated. Some curves according to different model 
predictions \cite{blha, bour, isla2, jll, ppp} are also drawn there. 
All of them fail to describe the data. We conclude that
namely this region becomes the Occam razor for all models. 

As we see in the Figures, various theoretical approaches have been attempted
for description of different regions of the differential
cross section. One by one we should name: 1. purely geometrical approach with
reference to the internal geometry of colliding hadrons, 2. the analogy to the
Fraunhofer diffraction, 3. the appeal to the electromagnetic and matter
density distributions inside hadrons, 4. the dynamical picture of Reggeon 
exchanges which is the most popular one with Pomeron playing the distinguished 
role, 6. the OCD-inspired models. Their detailed review is given in
\cite{ufnel}. Here, we restrict ourselves by several recent developments
not included or discussed there very briefly.

\section{Proton structure}

Since old days, it was clear that hadrons possess some internal structure.
Namely this we discussed with Pomeranchuk after I proposed the one-pion
exchange model of inelastic interactions \cite{dche}. It was 
treated as describing the peripheral interactions of hadrons. The deeper inside 
the hadron, the more pions should be exchanged and more dense populated should 
be the proton. Since then, the principal features have not been changed 
with pions replaced by partons (quarks and gluons) even though pions as 
a chiral anomaly play the distinguished role.

Early attempts to consider elastic scattering of hadrons also stemmed from 
the analogous simple geometrical treatment of their internal structure 
\cite{kris, chou, isl, cwu}. Starting from simplified pictures, one tried
to fit the elastic differential cross section. However, recent fits of LHC data 
failed outside the diffraction cone as demonstrated above. Thus these hypotheses
were not precise enough.

At the same time, one can get the direct knowledge about the proton structure
important for {\it inelasic} collisions using elastic scattering data.
The impact of the proton structure on inelastic processes can be viewed 
from the overlap function defined by the unitarity condition in 
$b$-representation (\ref{unib})
for elastic scattering amplitudes. It has been directly computed
\cite{dnec1} from experimental data of the TOTEM collaboration
for pp-scattering at 7 TeV and shown in Fig. 3.
\begin{figure}
\includegraphics[width=\textwidth, height=6.2cm]{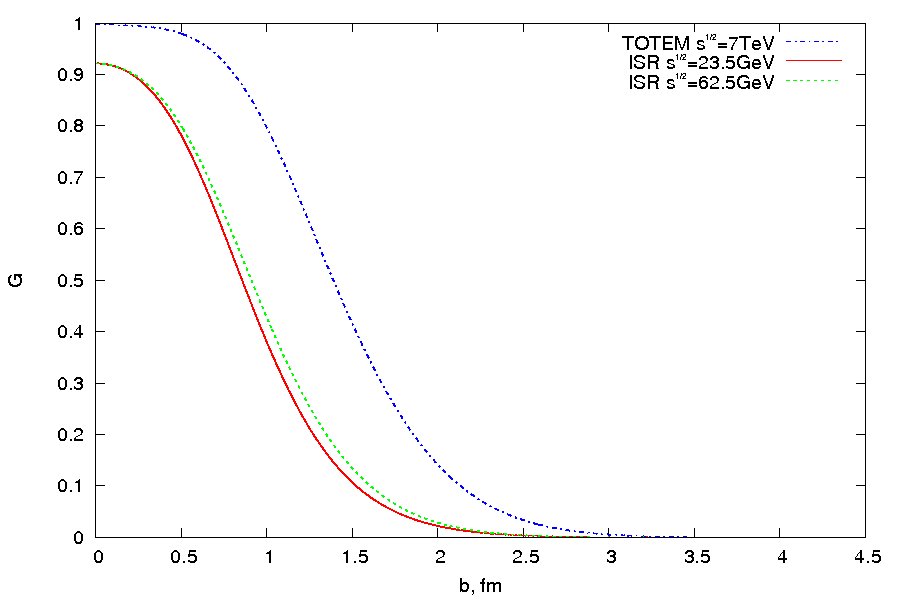}

Fig. 3. The overlap function $G(s,b)=4F(s,b)\leq 1$ at 7 TeV (upper curve)
\cite{dnec1} compared to those at ISR energies 23.5 GeV and 62.5 GeV. 
\end{figure}
The similar shape was obtained (see Fig. 4 in \cite{fsw})
assuming the Gaussian profile of the elastic contribution $h(s,b)$. Both shapes 
show the pattern with rather flat shoulder close to 1 (i.e. to complete
blackness) at small impact parameters $b$ and
subsequent quite steep fall-off. Attempts to fit it by a single Gaussian fail.

In view of the supposed proton substructure with a darker and stepwise
kernel surrounded by a more transparent cloud of partons it is reasonable
to attempt the fit with a stepwise behavior of the Gaussian exponential like
\begin{equation}
\ln \frac {F(s,0)}{F(s,b)}=\frac {b^2}{a[1-\frac {2}{\pi }\arctan \frac {b-b_0}
{\lambda }]}.
\end{equation}
The fit reveals quite strong separation of the two regions at 
$b_0 \approx 0.3$ fm with a width of the transition region 
$\lambda \approx 0.1$ fm. According to Eq. (5) the exponential becomes three
times larger in the narrow strip between the borders of the transition
region $b_0-\lambda $ and $b_0+\lambda $.
The region $b<b_0$ is completely black while
at $b>b_0$ it becomes more transparent. However when compared to ISR results
\cite{dnec1} its blackness increases with energy, especially at rather large
impact parameters about 1 fm as seen in Fig. 4.
\begin{figure}
\includegraphics[width=\textwidth, height=6.2cm]{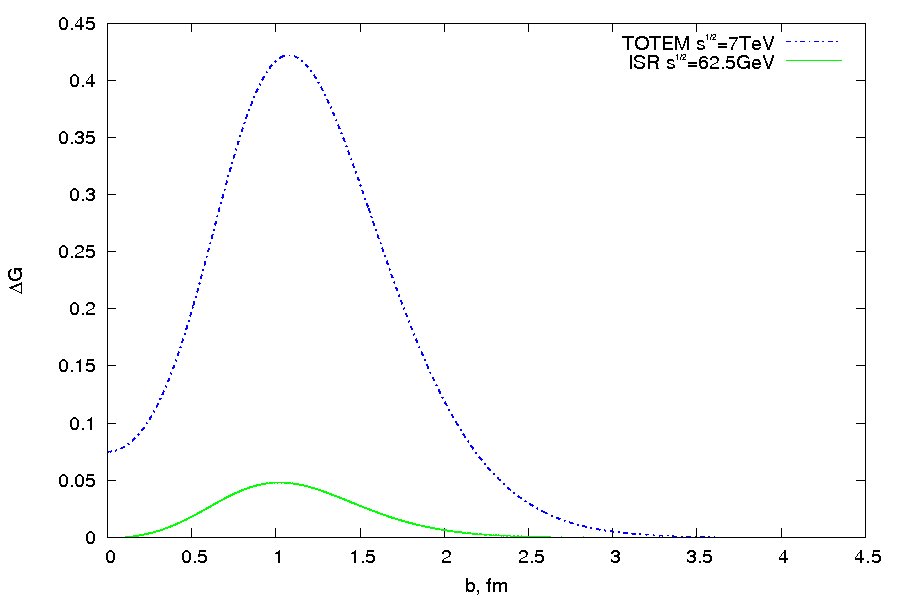}

Fig. 4. The difference between the overlap functions. Dash-dotted curve is for 
7 TeV and 23.5 GeV energies, solid curve is for 62.5 GeV and 23.5 GeV energies.
Conclusion: The parton density at the periphery increases strongly with energy
increase!
\end{figure}
These peculiar features have been used for description \cite{ads} of the CMS 
data \cite{022} about jet production at high multiplicities. The increased 
role of central pp-collisions with small impact parameters for events
triggered by jets was demonstrated. The important 
conclusion of this analysis is that the perturbative QCD can be applied to
these processes only at high enough transverse momenta of jets exceeding
7-8 GeV.

\section{Scaling laws}

Since long ago, it was discussed
\cite{akm71, ddd} a possibility that the differential cross sections might be
described as functions of a single scaling variable representing a definite 
combination of energy and transferred momentum. No rigorous proof of this
assumption has been proposed. This property was recently obtained \cite{drad}
from the solution of the partial differential equation for the imaginary part
${\rm Im}A(s,t)$ of the elastic scattering amplitude. The equation has been 
derived by equating the two expressions for the ratio of the real to imaginary
parts of the amplitude $\rho (s,t)$. They were known from the local dispersion 
relations (10) \cite{gmig, sukha, fkol} with the $s$-derivative and from the
linear $t$-approximation \cite{akm71, mar1} with the $t$-derivative (for more
details see \cite{drad}).

Therefrom the following partial differential equation is valid
\begin{equation}
p-f(x)q=1+f(x),
\label{partial}
\end{equation}
where $p=\partial u/\partial x ; \; q=\partial u/\partial y; \;
u=\ln {\rm Im}A(s,t); \; f(x)=2\rho (s,0)/\pi \approx d\ln \sigma _t/dx ; 
\; x=\ln s; \; y=\ln \vert t\vert ; \; \sigma _t$ is the total cross section. 
The variables $s$ and $\vert t\vert $ should be considered as scaled by the 
corresponding constant factors $s_0^{-1}$ and $\vert t_0\vert ^{-1}$.

The general solution of Eq. (\ref{partial}) reveals the scaling law
\begin{equation}
\frac {t}{s}{\rm Im}A(s,t)=\phi (t\sigma _t). 
\label{scal}
\end{equation}

For the differential cross section it looks like
\begin{equation}
t^2 d\sigma /dt=\phi ^2(t\sigma _t),
\label{sl1}
\end{equation}
if the real part of the amplitude is neglected compared to the imaginary part.
Thus the scaling law is predicted not for the differential cross section itself
but for its product to $t^2$.
Let us note that the often used ratio (see, e.g., \cite{bddd}) of $d\sigma /dt$
to $d\sigma /dt\vert _{t=0}\propto \sigma _t^2$ is also a scaling function.
However, the expression (\ref{sl1}) is more suitable for comparison with 
experiment.

The scaling law with the $t\sigma _t$-scale is known as the geometrical scaling.

\begin{figure}[h]
 \includegraphics[width=\textwidth, height=6cm]{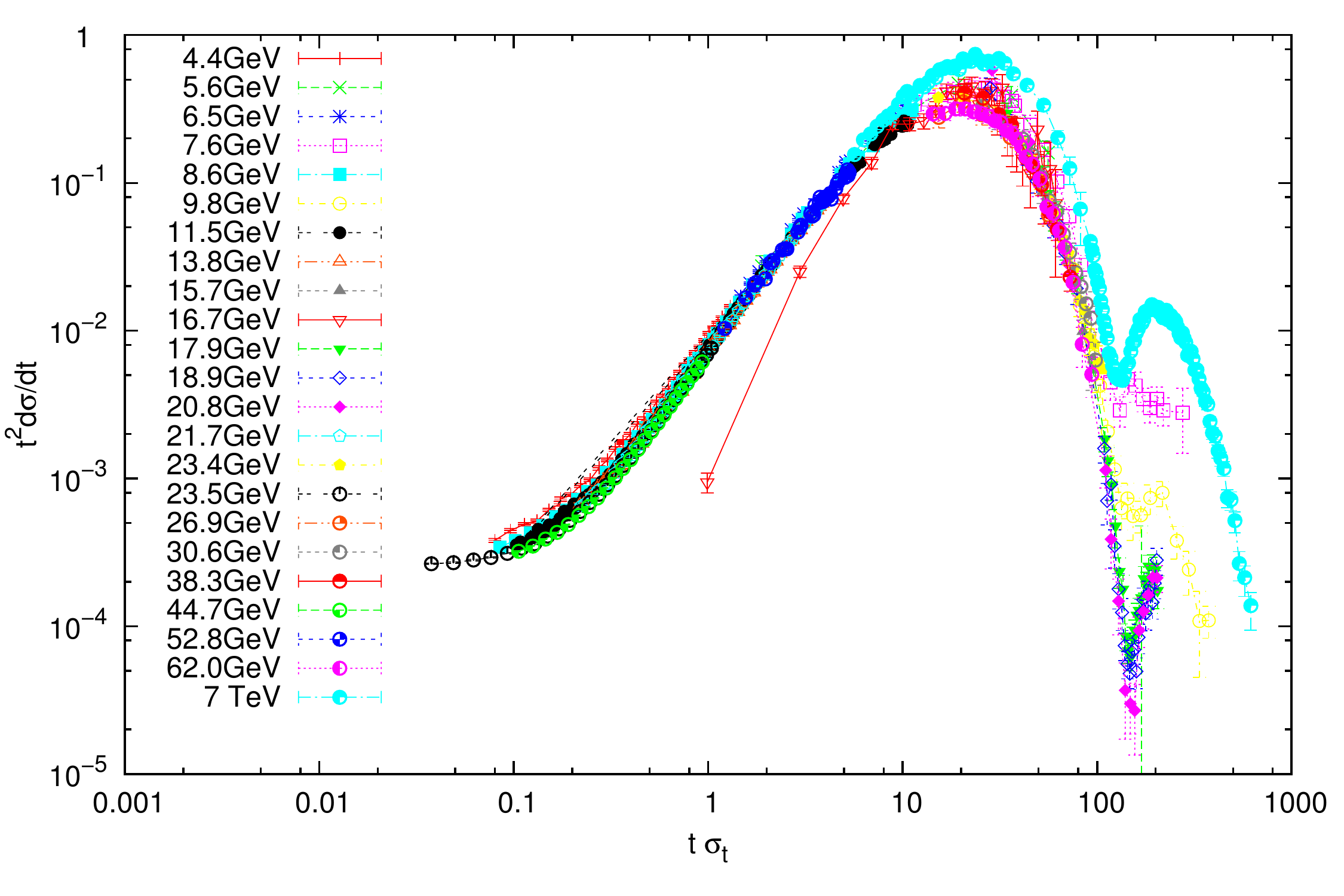}
Fig. 5. Violation of geometrical scaling at LHC energies.
 \label{fit_sigma_tot}
\end{figure}

In Fig. 5, we plot $t^2d\sigma /dt$ for pp-scattering at energies $\sqrt s$
from 4.4 GeV to 7 TeV as functions of $t\sigma _t$ with $\sigma _t$ provided by
the corresponding experiment. Rather reasonable scaling is observed in the
diffraction cone except the TOTEM data at 7 TeV. Thus the simple geometrical 
scaling is not fulfilled at high energies even at low transferred momenta. 
To restore some approximate scaling one should replace the variable $t\sigma _t$ 
by $t^a\sigma _t$ and plot $t^{2a}d\sigma /dt$ with $a\approx 1.2$ as shown 
in \cite{dnsc}. However, this is not a simple geometric scaling anymore. 

\section{Orear region and real part at any $t$}

The theoretical explanation of the new regime of exponential decrease of the
differential cross section beyond the diffraction cone with angles based on 
consequences of the unitarity condition in the $s$-channel has been proposed in Refs
\cite{anddre, anddre1}. The careful fit to experimental data showed good 
quantitative agreement with experiment \cite{adg}. Nowadays the same approach
helped explain the TOTEM findings \cite{dnec}.
 
We consider the lefthand side and the integral term $I_2$ in the unitarity 
condition (\ref{unit}) at the angles $\theta $ outside the diffraction peak.
Because of the sharp fall off of the amplitude with angle, the 
principal contribution to the integral arises from a narrow region around the
line $\theta _1 +\theta _2 \approx \theta $. Therefore one of the amplitudes
should be inserted at small angles within the cone as a Gaussian while another 
one is kept at angles outside it. Integrating over one of the angles the 
linear integral equation is obtained:        
\begin{equation}
{\rm Im}A(p,\theta )=\frac {p\sigma _t}{4\pi \sqrt {2\pi B}}\int _{-\infty }
^{+\infty }d\theta _1 e^{-Bp^2(\theta -\theta _1)^2/2} f_{\rho }
{\rm Im}A(p,\theta _1)+F(p,\theta ),
\label{linear}
\end{equation}
where $f_{\rho }=1+\rho _0\rho (\theta _1) $.

It can be solved analytically (for more details see \cite{anddre, anddre1})
with two assumptions that the role of the overlap function $F(p,\theta )$ is 
negligible outside the diffraction cone and the function $f_{\rho }$ may be
approximated by a constant, i.e. $\rho (\theta _1)=\rho _l$=const. 

It is esay to check that the eigensolution of this equation is
\begin{equation}
{\rm Im} A(p,\theta )=C_0\exp \left (-\sqrt 
{2B\ln \frac {Z}{f_{\rho }}}p\theta \right )+\sum _{n=1}^{\infty }C_n
e^{-({\rm Re }b_n)p\theta } \cos (\vert {\rm Im }b_n\vert p\theta-\phi _n)
\label{solut}
\end{equation}
with
\begin{equation}
b_n\approx \sqrt {2\pi B\vert n\vert}(1+i{\rm sign }n) \;\;\;\;\;\;\; n=\pm 1, \pm 2, ...
\label{bn}
\end{equation}
This expression contains the exponentially decreasing with $\theta $ (or 
$\sqrt {\vert t \vert }$) term (Orear regime!) with imposed on it  
oscillations strongly damped by their own exponential factors. These oscillating
terms are responsible for the dip. The exponential in Eq. (\ref{solut}) is well 
defined. It contains the value of $Z$ in the logarithm  
becoming very sensitive to $\rho $ when $Z$ approaches 1 (see Table 2). 
The comparison with LHC data has
shown that this ratio must be negative and quite large (about -2) in this
region. Most of the widely used models do not predict such values. Moreover
many of them get it positive. This follows from the equal numbers of zeros of 
real and imaginary parts. Only those models with odd sum of this number
can succeed in getting negative $\rho (s,t)$. 
The unitarity condition does not ask for a zero of the imaginary part to fit
the dip as the models do but ascribes it to the damped oscillations contained
in the solution of the equation. This discrepancy is not resolved yet. 
No correspondance between the two approaches has yet been established. Some 
equations for the ratio have been derived and solved. They favor the variable 
sign of it. However the problem asks for further investigation. 

\section{Conclusions}

There are new exciting findings at LHC. They pose serious problems for
theoreticians. I am sure that all of them would be of great interest to
I.Ya. Pomeranchuk.

\end{document}